# The speed of gravity revisited

Michael Ibison, Harold E. Puthoff, Scott R. Little
*Institute for Advanced Studies at Austin*
*4030 Braker Lane West, Suite 300, Austin, TX 78759, USA*
ibison@ntr.net, puthoff@aol.com, little@eden.com

## Abstract

Recently Van Flandern concluded from astrophysical data that gravity propagates faster than light. We demonstrate that the data can be explained by current theory that does not permit superluminal speeds. We explain the origin of apparently instantaneous connections, first within EM, and then within strong-field GR.

## Introduction

Van Flandern [1] draws attention to astrophysical data that apparently support the conclusion that gravitational influences propagate at superluminal speeds. His main argument, in general terms, is that light propagation from a star is not collinear with its gravitational force. He correctly infers that this is because light suffers aberration, whereas gravity does not. The absence of the latter motivates his conclusion.

In the first section we discuss the related example of the electric field of a uniformly moving charge source. We show how this field, though composed of entirely retarded influences obeying Maxwell's equations, will accelerate a remote test charge towards the instantaneous position of the source. In the second section we discuss the analogous problem in GR. On the basis of the similarities between EM and GR, we argue that a similar result can be anticipated, for which the detailed calculation is given in an appendix. That section is concluded with an illustration of the application of the theory to astrophysical data offered by Van Flandern.

## Electromagnetism

*Electric field of a uniformly moving charge*

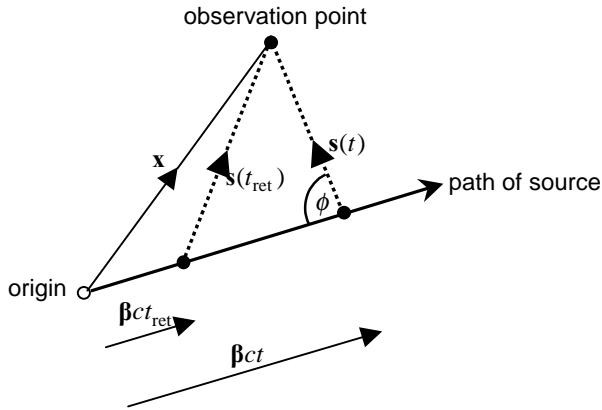

**Figure 1**

Quantities used to analyze a uniformly moving charge source.

With reference to Fig. 1, let a source charge be in uniform motion traveling with velocity $c\boldsymbol{\beta}$, and initially at the origin. It generates an electric field at **x** (see for example [2]):

$$\mathbf{E}(\mathbf{x},t) = e \frac{\mathbf{s}(t_{\text{ret}}) - s(t_{\text{ret}})\boldsymbol{\beta}}{\gamma^2 (s(t_{\text{ret}}) - \mathbf{s}(t_{\text{ret}}) \cdot \boldsymbol{\beta})^3}, \quad (1)$$

$\mathbf{s}(t)$ is the vector from the current position of the source to the test charge at **x**:

$$\mathbf{s}(t) = \mathbf{x} - \boldsymbol{\beta}ct. \quad (2)$$

The subscript *ret* in the above indicates that the electric field depends on the retarded position of the source, with $t_{\text{ret}}$ given by the solution of

$$s(t_{\text{ret}}) = c(t - t_{\text{ret}}). \tag{3}$$

That is, the current field at time *t* depends on the position of the source at a previous time, where the delay is equal to the time it takes light to traverse that distance (from the historical position of the source). We draw attention especially to the presence of two retarded terms in the numerator, giving rise to the interpretation of the final field as the result of two different, retarded, fields, $\mathbf{E}(\mathbf{x},t) = \mathbf{E}_1(\mathbf{x},t) + \mathbf{E}_2(\mathbf{x},t)$:

$$\mathbf{E}_1(\mathbf{x},t) = e \frac{\mathbf{s}(t_{\text{ret}})}{\gamma^2 \left(s(t_{\text{ret}}) - \mathbf{s}(t_{\text{ret}}) \cdot \boldsymbol{\beta}\right)^3}$$

$$\mathbf{E}_2(\mathbf{x},t) = -e \frac{s(t_{\text{ret}})\boldsymbol{\beta}}{\gamma^2 \left(s(t_{\text{ret}}) - \mathbf{s}(t_{\text{ret}}) \cdot \boldsymbol{\beta}\right)^3} \tag{4}$$

If only $\mathbf{E}_1(\mathbf{x},t)$ is retained, it is easy to see that the electric force on a test charge would be oriented towards the historical position of the source. I.E. it would be aberrated. One might then be able to take the view that the force is due to a flux of particles emitted at the speed of light, exchanging momentum with the test charge.

*Apparent cancellation of retarded effects*

Since $\beta < 1$, the second term may be regarded as a correction to the first. As pointed out in [2], [3], and [4], its magnitude and direction are exactly that required to cancel the aberration as we now show. Combining Eqs. (2) and (3), the numerator of Eq. (1) can be written:

$$\mathbf{s}(t_{\text{ret}}) - s(t_{\text{ret}})\boldsymbol{\beta} = \mathbf{x} - ct_{\text{ret}}\boldsymbol{\beta} - c(t - t_{\text{ret}})\boldsymbol{\beta} = \mathbf{x} - ct\boldsymbol{\beta} = \mathbf{s}(t) \tag{5}$$

which is sufficient to prove the claim. Note that this result depends on the fact that the source is in uniform motion. To complete the transformation we will also compute the denominator in the new co-ordinates. Using Eq. (5), we have

$$s(t_{\text{ret}}) - \mathbf{s}(t_{\text{ret}}) \cdot \boldsymbol{\beta} = (1 - \beta^2) s(t_{\text{ret}}) - \mathbf{s}(t) \cdot \boldsymbol{\beta} = s(t)\left((1 - \beta^2)\rho - \beta \cos(\phi)\right) \tag{6}$$

where $\phi$ is the angle between the trajectory of the source and the instantaneous line of sight (see Fig. 1), and we have defined $\rho \equiv s(t_{\text{ret}})/s(t)$. This may also be computed from Eq. (5) after re-arranging and squaring:

$$s^2(t_{ret}) = (\mathbf{s}(t) - s(t_{ret})\boldsymbol{\beta})^2$$
$$\Rightarrow (1 - \beta^2)\rho^2 - 2\rho\beta\cos(\phi) - 1 = 0 \tag{7}$$
$$\Rightarrow (1 - \beta^2)\rho = \beta\cos(\phi) + \sqrt{1 - \beta^2 \sin^2(\phi)}$$

(since by definition $\rho$ must always be positive, the positive root is required). Substitution of this result into Eq. (6) gives

$$s(t_{\text{ret}}) - \mathbf{s}(t_{\text{ret}}) \cdot \boldsymbol{\beta} = s(t)\sqrt{1 - \beta^2 \sin^2(\phi)}. \tag{8}$$

Therefore, using Eqs. (5) and (8), the electric field Eq. (1) in the unretarded coordinates is

$$\mathbf{E}(\mathbf{x},t) = e\frac{\hat{\mathbf{s}}(t)}{s^2(t)} \times \left\{ \frac{1 - \beta^2}{\left(1 - \beta^2 \sin^2(\phi)\right)^{3/2}} \right\} \tag{9}$$

This result proves the claim above that the electric field from a uniformly moving source is not aberrated. The force on a test charge is directed towards the instantaneous – not the retarded - position of the source. The factor in braces affects the magnitude, but not the direction, of the field. When the source is viewed from a line of sight perpendicular to its trajectory, then the amplitude of the field is greater than that of the static charge by a factor $\gamma$. But viewed from a line of sight along the trajectory, the amplitude is reduced by a factor $\gamma^2$. The effects of both non-retardation and the angular dependence of the strength of the force due to a moving source are illustrated in Fig. 2. Therein, the density of the field lines signifies the strength of the field, and the direction of the field lines is that of the force experienced by a stationary test charge.

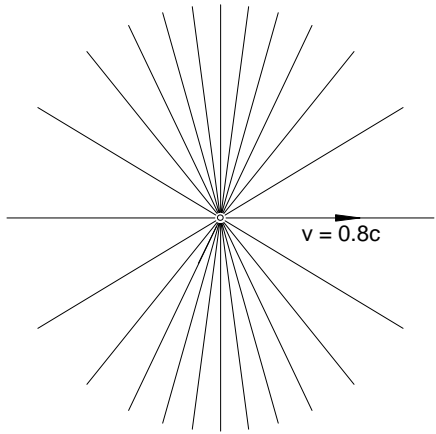

**Figure 2**

Field lines of a uniformly moving charge source.

The fact that there are two mathematical presentations of the same result, Eqs. (1) and (9), can give rise to two different interpretations of the same phenomenon. But since they are built upon the same theory, these two interpretations are not experimentally distinguishable. Motivated by Eq. (1), one can insist that all influences travel at light speed, and that the final direction of the force toward the unretarded position is the result of a fortuitous cancellation between the two retarded influences identified in Eq. (4). Alternatively, Eq. (9) may be used to justify the viewpoint that there is no propagation, since nowhere do retarded co-ordinates appear, and that the result is a manifestation of an important principle, and presumably could therefore have been derived more directly [5].

*Propagation*

When the charge source is accelerated, Eq. (2) does not hold, invalidating the steps leading to Eq. (9). Further, the electric field acquires additional terms that depend on the acceleration directly. Altogether, this means that the force on a test charge is not, in general, directed toward an accelerating source, and the effects of retardation are then readily apparent.

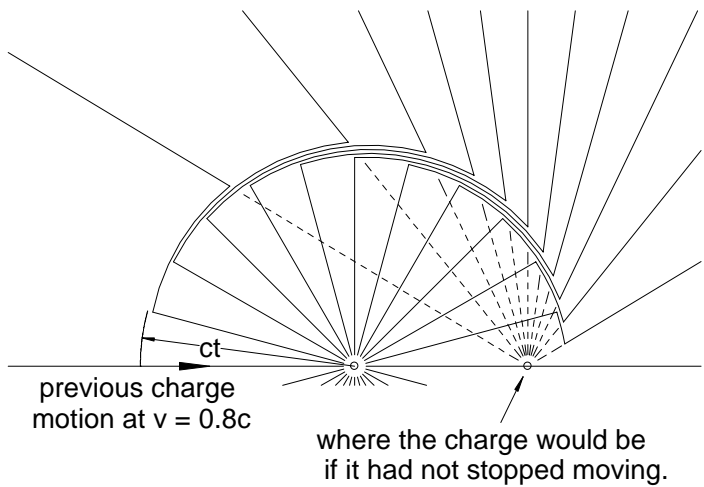

**Figure 3**

Propagation of a disturbance to the field lines when a uniformly moving charge is abruptly brought to rest.

Observable propagation delays are illustrated by the example in Fig. 3 wherein a charge initially in uniform motion is abruptly brought to rest. Notice that the field lines are updated by a shockwave traveling at speed $c$. Within a radius $r < ct$, the charge sprouts new field lines. Beyond that radius ($r > ct$), the field lines appear to emanate from a source location that the charge does not occupy. The shockwave at $r = ct$ carries energy and momentum, and is a manifestation of radiation. More generally: curved field lines and radiation are generated and propagate whenever there is a departure from uniform motion (of either source or observer). Otherwise, if the motion has been uniform for some period $\tau$, say, then the field lines within the radius $r < c\tau$ are straight and terminate on the instantaneous position of the source.

### *Role of the potentials in the cancellation of retarded effects*

There is nothing in GR that corresponds exactly to the electric field of EM. But there *is* a close correspondence between the scalar potentials of both theories [6]. Therefore, to determine the degree to which the non-propagating aspect of EM applies to GR, we need to determine the relative importance of the scalar potential - versus the vector potential - in establishing this result. In terms of the potentials, the electric field is

$$\mathbf{E}(\mathbf{x},t) = -\nabla\phi - \frac{1}{c}\frac{\partial \mathbf{A}}{\partial t}. \tag{10}$$

Recalling the decomposition of the electric field as expressed in Eq. (4), one could be forgiven for thinking that perhaps $\mathbf{E}_1(\mathbf{x},t)$ and $\mathbf{E}_2(\mathbf{x},t)$ correspond respectively to $-\nabla\phi$ and $-(1/c)\partial\mathbf{A}/\partial t$. However, this turns out not to be the case as can be seen from the following argument. For a point source, the Liénard-Wiechert potentials are

$$\{\phi, \mathbf{A}\} = \frac{e\{1, \boldsymbol{\beta}\}}{\sigma}, \tag{11}$$

where the dependence of the denominator on $\mathbf{x}$ and the *present* time can be found from Eq. (8):

$$\sigma \equiv s(t_{\text{ret}}) - \mathbf{s}(t_{\text{ret}}) \cdot \boldsymbol{\beta} = s(t)\sqrt{1 - \beta^2 \sin^2(\phi)} = \sqrt{s^2(t) - (\boldsymbol{\beta} \times \mathbf{s})^2} = \sqrt{(\mathbf{x} - ct\boldsymbol{\beta})^2 - (\boldsymbol{\beta} \times \mathbf{x})^2}. \tag{12}$$

Given the co-occurrence of $\beta$ with $ct$ in this expression, and the dependence of $\mathbf{A}$ on $\boldsymbol{\beta}$ as given in Eq. (11), it is readily deduced that the contribution to the electric field from the time derivative of the vector potential is of order $\beta^2$ or higher and cannot therefore play a role at low velocities. It follows that the gradient of the scalar potential alone gives the correct electric field up to and including terms of order $\boldsymbol{\beta}$. This means that, to that order, the direction of the electric force towards the instantaneous position of the source may be regarded entirely as a property of the form of the scalar potential. I.E. from Eq. (12), to order $\boldsymbol{\beta}$: $\sigma = s(t)$, and so $\phi = -e/|\mathbf{x} - ct\boldsymbol{\beta}| + \mathbf{O}(\beta^2)$. Thus, to this order, one finds that the Coulomb potential is as if not retarded, and

$$\mathbf{E}(\mathbf{x},t) = -e\nabla\frac{1}{|\mathbf{x} - ct\boldsymbol{\beta}|} + \mathbf{O}(\beta^2) = e\frac{\hat{\mathbf{s}}(t)}{s^2(t)} + \mathbf{O}(\beta^2). \tag{13}$$

Hence, a shorthand description valid only to order $\boldsymbol{\beta}$ is that the <u>Coulomb</u> potential is not retarded, from which it follows that the force is directed towards the instantaneous position of the source. The more accurate description, as shown in detail above, is that the <u>scalar</u> potential computed from solving Maxwell's equations *is* retarded. But it does not have the form one might naively expect; it is not a retarded version of the Coulomb potential. I.E. $\phi \neq -e/|\mathbf{x} - ct_{\text{ret}}\boldsymbol{\beta}|$, since this is <u>not</u> a solution of Maxwell's equations. Instead, $\phi = -e/|\mathbf{x} - ct\boldsymbol{\beta}| + \mathbf{O}(\beta^2)$ is the solution of Maxwell's equations. Comparing Eqs. (9) and (13), it follows that at low velocities, the vector potential plays no role in establishing the non-propagating electric force.

## Gravitation

*Acceleration due to a moving source mass*

A form analogous to Eq. (10) can be found for the proper acceleration attributable to gravity [7]. Unlike the EM case however, there is no exact cancellation of the effects of retardation to all orders of $\beta$, but only cancellation up to terms linear in $\beta$. In the linearized weak-field limit, cancellation to this order can be anticipated from the EM case on the basis of the correspondence between the gravitational and electromagnetic scalar potentials. Both obey wave equations, and under the circumstances of interest here, the sources of both may be idealized as structureless points. Hence both gravitational and electromagnetic scalar potentials fall off as $1/r$ when the sources are static. But the correspondence cannot be perfect because mass-energy – unlike charge – is not independent of velocity. This causes the two theories to diverge at the order of $\beta^2$. From these observations, one can infer that GR should be in agreement with EM at least up to terms linear in $\beta$, and, with reference to the above discussion of Eq. (13), that this correspondence can be established solely from consideration of the gravitational scalar potential. In other words, without any analysis, one can anticipate that Newton's law of gravity - unretarded – is accurate up to terms linear in $\beta$. The details are worked out in the appendix, where it is established that the proper acceleration in the weak field limit is

$$\frac{d^2\mathbf{x}}{d\tau^2} \approx -\frac{GM_0\hat{\mathbf{s}}(t)}{s^2(t)} + \mathbf{O}(\beta^2). \tag{A18}$$

It turns out that not only the linearized weak field limit, but also the full moving Schwarzschild solution to the Einstein equations gives an acceleration towards the instantaneous position of the source (mass $M_0$), correct to order $\beta$. From the appendix, the result is

$$\frac{d^2\mathbf{x}}{d\tau^2} = -\frac{GM_0\hat{\mathbf{s}}(t)}{\left(1-\frac{GM_0}{2c^2s(t)}\right)\left(1+\frac{GM_0}{2c^2s(t)}\right)^5 s^2(t)} + \mathbf{O}(\beta^2). \tag{A17}$$

The steps leading to these solutions involve computing derivatives of time-retarded potentials, just as for the EM case. Yet the result to this order is the same: the $\mathbf{r}/r^3$ force law remains unchanged and unretarded if the source is moving.

*Solar Eclipse*

Van Flandern offers several astrophysical situations in support of his conclusion that the speed of gravity $\gg c$. Rather than attempt to address all of them in this letter, we have chosen to discuss just one in detail. Hopefully it will be clear how the analysis can be extended to cover his other examples.

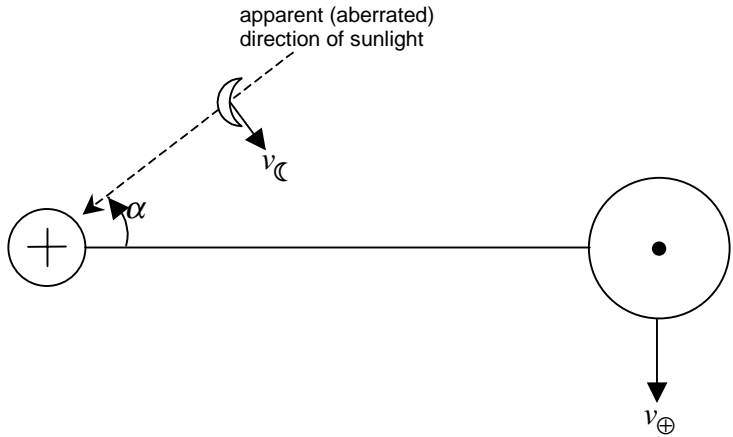

**Figure 4a**

Situation at the time of solar eclipse in earth-centered coordinates.

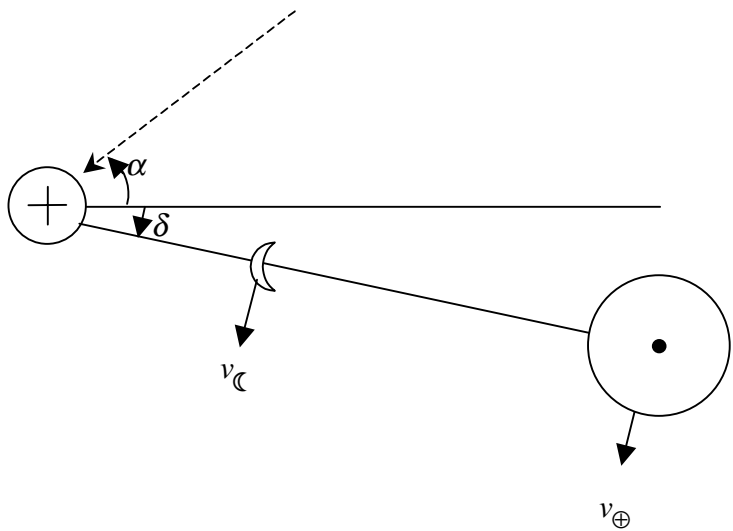

**Figure 4b**

Situation at the time of conjunction of sun, moon, and earth, in earth-centered coordinates.

The situation during a solar eclipse is depicted in Fig. 4, wherein all motion is relative to the earth. Van Flandern points out that there are two events that happen around the same time. The first event is when the moon crosses the path of radiation from the sun to the earth, as depicted in Fig. 4a. This is the time that the sun is eclipsed as seen from the earth. But the line of sight of the sun is not the same as the line from the earth to the current position of the sun because of the time delay of light propagation. The angle $\alpha$ between the two - the aberration angle - is approximately given by the ratio $v_\oplus/c = \omega_\odot t_{\odot\to\oplus}$ where $t_{\odot\to\oplus} = 499$ seconds is the light-time from sun to earth, and $\omega_\odot = 2\pi$ (years)$^{-1}$ is the angular frequency attributable to the sun in an earth-centered coordinate system. Sometime $\Delta t$ later the sun, moon and earth line up (conjunct), as depicted in Fig. 4b. The time of this second event can be calculated by noting that the time it takes for the moon to traverse the angle an angle $\alpha + \delta$ is equal to the time it takes the sun to traverse $\delta$:

$$\Delta t = \frac{\delta}{\omega_\odot} = \frac{\alpha + \delta}{\omega_\leftmoon} \qquad (14)$$

This gives $\delta = \alpha\omega_\odot/(\omega_\text{☾} - \omega_\odot)$ and therefore that

$$\Delta t = \frac{\alpha}{\omega_\text{☾} - \omega_\odot} = \frac{t_{\odot \to \oplus}}{\omega_\text{☾}/\omega_\odot - 1}. \tag{15}$$

Using the ratio of the two frequencies = 13.4 (the number of lunar months per year), the time interval between the two events is 40.2 seconds. Van Flandern points out that observations indicate that the acceleration of the earth toward the sun is maximal at around 40 seconds after the sun is eclipsed and therefore this is the time of maximal gravitational attraction. From this he infers that the speed of gravity must be much faster than light.

The theoretical explanation of this observation is as follows. Sunlight is electromagnetic radiation due (classically) to the acceleration of charges in the sun. Such radiation suffers a propagation delay as illustrated in Fig. 3, and is therefore aberrated. By contrast, gravitational radiation due to the earth's acceleration is negligible, leaving only the linear aspect of the relative sun-earth motion to consider. Since $v_\oplus/c \ll 1$, the results of the previous section apply, and one concludes that the gravitational force on the earth is directed towards the instantaneous position of the sun; i.e. is not aberrated.

*Generalization*

From this example we can draw a more general conclusion that applies to Van Flandern's other data. Noticing that separated aggregates of matter tend to be electrically neutral, one concludes that the predominant electromagnetic force in astrophysics will be due to radiation (in the far-field) - the non-radiative (near-field) force between charged bodies predominates only at a very small scale. In contrast, because gravity is always attractive, the non-radiative force does not disappear between aggregates of matter. The force of gravitational radiation is comparatively negligible, and therefore the predominant gravitational force in astrophysics will be non-radiative (near-field). Hence, though the qualities of propagation and non-propagation of the far and near fields respectively are the same for both EM and GR, the dominance of radiative EM interactions and non-radiative GR interactions explains the asymmetry of influences in this and other examples.

## Conclusion

Van Flandern is correct in his observation that gravitational attraction is directed towards the instantaneous (unretarded) position of a moving body. We have shown that this fact can be explained without a revision of physics to include superluminal propagation.

# Appendix
# Gravitational force on a static test particle due to a moving source

This appendix is in two sections. In the first section a 3-vector form of the geodesic equation is presented for a test mass initially at rest and solely under the influence of a gravitational field via the metric. In the second section the problem is solved for a gravitational field of arbitrary strength using the Schwarzschild metric for a moving source.

*3-vector form of geodesic equation*

The general equation of motion for an infinitesimally-sized, infinitesimally-massive (test) particle with only gravitational forces acting is

$$\frac{d}{d\tau}\left(g_{\alpha\beta}\frac{dx^\beta}{d\tau}\right) = \frac{1}{2}\frac{\partial g_{\beta\gamma}}{\partial x^\alpha}\frac{dx^\beta}{d\tau}\frac{dx^\gamma}{d\tau} . \tag{A1}$$

Going to 3D notation, let

$$\mathbf{G} = \begin{pmatrix} g_{11} & g_{12} & g_{13} \\ g_{21} & g_{22} & g_{23} \\ g_{31} & g_{32} & g_{33} \end{pmatrix}, \quad \mathbf{g} = \begin{pmatrix} g_{10} \\ g_{20} \\ g_{30} \end{pmatrix} \tag{A2}$$

and let the velocity of the test particle, $d\mathbf{x}/dt$, be zero. Then the 3-vector part of Eq. (A1) gives

$$\frac{d}{d\tau}\left(c\Gamma\mathbf{g} + \mathbf{G}\frac{d\mathbf{x}}{d\tau}\right) = \frac{c^2\Gamma^2}{2}\nabla g_{00} , \tag{A3}$$

where, when the velocity of the test mass is zero, the coordinate time is related to the proper time by $\Gamma \equiv dt/d\tau = 1/\sqrt{-g_{00}}$. With this substitution, Eq. (A3) may be written

$$m_0 \frac{d^2\mathbf{x}}{d\tau^2} = -\frac{m_0 c^2}{\sqrt{-g_{00}}}\mathbf{G}^{-1}\left(\nabla\sqrt{-g_{00}} + \frac{1}{c}\frac{\partial}{\partial t}\frac{\mathbf{g}}{\sqrt{-g_{00}}}\right). \tag{A4}$$

Expressed in this way, the geodesic equation takes on the appearance of Newton's second law, where the expression on the right-hand side is a driving force attributable to the gravitational field. Møller [7] noted the similarity between this expression for the force, and that of the electric component of the Lorentz force,

i.e. $$e\mathbf{E} = -e\left(\nabla\phi + \frac{1}{c}\frac{\partial \mathbf{A}}{\partial t}\right) \tag{A5}$$

and accordingly defined a gravitational scalar and vector potential. Forward [8] subsequently pursued this idea by identifying a magnetic-like component of force on a *moving* test body. However, as admitted by Forward, the correspondence between GR and EM is only approximate. In EM, conservation of electric charge, expressed as a vanishing 4-divergence in the source densities, $\partial^\mu j_\mu = 0$, may be translated into a vanishing 4-divergence of the potentials $\partial^\mu A_\mu = 0$ By contrast, in GR, the vanishing 4-divergence in the source densities, $\partial^\mu T_{\mu\nu} = 0$, does not translate into a GR equivalent of (A5), and therefore

$$\nabla\frac{\mathbf{g}}{\sqrt{-g_{00}}} + \frac{1}{c}\frac{\partial\sqrt{-g_{00}}}{\partial t} \neq 0 . \tag{A6}$$

Consequently care should be exercised in attempting to draw EM-inspired conclusions from Eq. (A4), even when the test body is static. For this reason, we do the analysis, rather than assume that the electromagnetic result (that the force on a uniformly moving test body points to the instantaneous position of the source) applies to gravity.

*Schwarzschild metric*

The full metric corresponding to the Schwarzschild solution for a moving mass has been given by [9]. The result is

$$g_{\alpha\beta}(x_\mu) = (1+X)^4 (\eta_{\alpha\beta} + u_\alpha u_\beta) - \left(\frac{1-X}{1+X}\right)^2 u_\alpha u_\beta \tag{A7}$$

where

$$X = \frac{\lambda}{\gamma\sigma}; \quad \lambda = \frac{GM_0}{2c^2} \tag{A8}$$

where $M_0$ is the rest mass of the actively gravitating body, and $\sigma$ is given by Eq. (12). One may easily extract $g_{00}$, **g**, and **G** from $g_{\alpha\beta}$:

$$g_{00} = -\gamma^2 \left(\left(\frac{1-X}{1+X}\right)^2 - (1+X)^4 \beta^2\right) \tag{A9}$$

and

$$\mathbf{g} = \gamma^2 \left((1+X)^4 - \left(\frac{1-X}{1+X}\right)^2\right)\boldsymbol{\beta} \tag{A10}$$

and

$$\mathbf{G} = (1+X)^4 \left(\mathbf{I} + \left(1 - \frac{(1-X)^2}{(1+X)^6}\right)\gamma^2 \boldsymbol{\beta}\boldsymbol{\beta}^T\right). \tag{A11}$$

These expressions may be inserted into the right hand side of Eq. (A4) to obtain an exact closed form expression for the force. However, we already know in advance that since the weak field force is not directed towards the source (if terms beyond those linear in $\beta^2$ are significant) then under the same conditions the exact (strong field) solution cannot be directed towards the source either. The only remaining interest then is to determine if the force is directed toward the source for all strengths of field up to and including linear terms in $\boldsymbol{\beta}$. Rewriting the equation of motion (A4) as

$$\frac{d^2\mathbf{x}}{d\tau^2} = -\frac{c^2}{2g_{00}} \mathbf{G}^{-1}\left(\nabla g_{00} - \frac{2}{c}\frac{\partial \mathbf{g}}{\partial t} + \frac{\mathbf{g}}{c}\frac{\partial}{\partial t}\log(-g_{00})\right) \tag{A12}$$

it is easy to show that the second two terms are of order $\beta^2$. First note that $X|_{\boldsymbol{\beta}=0} = \lambda/x$ and therefore that $\partial X|_{\boldsymbol{\beta}=0}/\partial t = 0$ which immediately gives (from (A10)) that $\partial \mathbf{g}/\partial t$ must be at least second order in $\beta$. Similarly, it follows that $\partial g_{00}/\partial t$ must be at least first order in $\beta$, from which it follows (again using (A10)) that $\frac{\mathbf{g}}{c}\frac{\partial}{\partial t}\log(-g_{00})$ must be at least second order. Further, since $\mathbf{G} = (1+X)^4\mathbf{I} + \mathbf{O}(\beta^2)$, then $\mathbf{G}^{-1} = (1+X)^{-4}\mathbf{I} + \mathbf{O}(\beta^2)$. Therefore, the equation of motion Eq. (A12) is

$$\frac{d^2\mathbf{x}}{d\tau^2} = -\frac{c^2}{2(1+X)^4}\nabla\log(g_{00}) + \mathbf{O}(\beta^2), \tag{A13}$$

i.e., for small velocities, the force depends only on the (gradient of the) scalar gravitational potential, and not on the 'gravitational vector potential'. To compute the gradient of $g_{00}$, first note that

$$g_{00} = -\left(\frac{1-X}{1+X}\right)^2 + \mathbf{O}(\beta^2) \tag{A14}$$

and also $\sigma = s(t)$ to order $\boldsymbol{\beta}$, so $X = \lambda/s(t) + O(\beta^2)$. Then

$$\nabla g_{00} = \frac{4(1-X)\nabla X}{(1+X)^3} + \mathbf{O}(\beta^2) = -\frac{4\lambda(1-X)\nabla s(t)}{(1+X)^3 s(t)^2} + \mathbf{O}(\beta^2) = -\frac{4\lambda(1-X)\mathbf{s}(t)}{(1+X)^3 s(t)^3} + \mathbf{O}(\beta^2) \tag{A15}$$

hence

$$\nabla \log(g_{00}) = \frac{4\lambda \mathbf{s}(t)}{(1-X^2)s(t)^3} + \mathbf{O}(\beta^2) = \frac{4\lambda \hat{\mathbf{s}}(t)}{(s^2(t)-\lambda^2)} + \mathbf{O}(\beta^2). \tag{A16}$$

Inserting Eqs. (A16) into (A13) gives

$$\frac{d^2\mathbf{x}}{d\tau^2} = -\frac{GM_0 \hat{\mathbf{s}}(t)}{\left(1-\frac{GM_0}{2c^2 s(t)}\right)\left(1+\frac{GM_0}{2c^2 s(t)}\right)^5 s^2(t)} + \mathbf{O}(\beta^2). \tag{A17}$$

In the weak field limit, the proper acceleration becomes

$$\frac{d^2\mathbf{x}}{d\tau^2} \to -\frac{GM_0 \hat{\mathbf{s}}(t)}{s^2(t)} + \mathbf{O}(\beta^2) \tag{A18}$$

which – as expected – has the same form as the EM result for a uniformly moving charge, Eq. (13). Note that the directions of the ordinary and the proper acceleration are the same since d$\mathbf{x}$/d$t$ = **0.**